\documentclass[%
 reprint,
 amsmath,amssymb,
 aps,
]{revtex4-2}

\usepackage{graphicx}
\usepackage{dcolumn}
\usepackage{bm}

\begin{document}

\preprint{APS/123-QED}

\title{Nuclear quantum effect on the elasticity of ice VII under pressure:\\ A path-integral molecular dynamics study}

\author{Jun Tsuchiya}
 \altaffiliation[Also at ]{Earth Life Science Institute (ELSI), Tokyo Institute of Technology.}
\affiliation{%
 Geodynamics Research Center, Ehime University\\
 2-5 Bunkyo-cho, Matsuyama Ehime 790-8577 Japan 
}%
\author{Motoyuki Shiga}
\affiliation{
Japan Atomic Energy Agency\\
 148-4 Kashiwanoha Campus, 178-4 Wakashiba, Kashiwa, Chiba, 277-0871, Japan}%
\author{Shinji Tsuneyuki}
\affiliation{%
 Department of Physics, The University of Tokyo
\\Hongo, Bunkyo-ku, Tokyo 113-0033, Japan}%

\author{Elizabeth C. Thompson}
\affiliation{%
 Department of Earth and Environmental Systems, The University of the South
\\735 University Avenue, Sewanee, TN 37383, USA}%

\date{\today}

\begin{abstract}
We investigate the effect of nuclear quantum effects (NQEs) of hydrogen atoms on the elasticity of ice VII at high pressure and ambient temperature conditions using ab initio path-integral molecular dynamics (PIMD) calculations. We find that the NQEs of hydrogen contributes to the transition of ice VII from a static disordered structure to a dynamically disordered structure at pressures exceeding 40 GPa. This transition is marked by a discontinuous increase of the elastic constants. Comparison of ab initio molecular dynamics and PIMD calculations reveal that NQEs increase the elastic constants of ice by about 20\% at 70 GPa and 300 K. 
\end{abstract}

\maketitle


\section{\label{sec:level1}Introduction\protect}
Determining the structures and physicochemical properties of water ice (H$_2$O) is of fundamental importance to the fields of physics, chemistry, and planetary science. Although numerous experimental and theoretical studies have been performed on the twenty known crystalline polymorphs of water ice \citep{hansen2021}, their physical and chemical properties are not yet fully understood. An intrinsic challenge in studying ices is that hydrogen is the lightest atom (Z = 1), and therefore hydrogen is extremely mobile, difficult to detect experimentally, and exhibits significant quantum effects.

The complexity and elusiveness of hydrogen is particularly evident when probing the behavior of ice at extreme pressures. As pressure is increased at ambient temperature, H$_2$O changes from ice VII to ice X. Ice VII is a cubic crystal structure with static orientational disorder of the H$_2$O molecules while maintaining the ice rule \citep{bernal1933}. This is possible because the oxygen atoms in ice VII are coordinated by four proton sites that are 50\% occupied, and the orientation of each H$_2$O molecule is dictated by which two of these four sites are occupied by hydrogen \citep{guthrie2019}. The ice VII structure, and therefore its physical properties, are driven by this hydrogen bond
network, which contains asymmetric hydrogen bonds \citep{tsuchiya2017}. In contrast, ice X is an ionic crystal with completely symmetric hydrogen bonds \citep{holzapfel1972, schweizer1984}. 

Broadly speaking, a transition in which the hydroxyl bonds ({O\textendash H}) and hydrogen bonds ({O$\cdot\cdot\cdot$H}) in a crystal structure have the same length and form a straight line (O\textendash H$\cdots$O $\approx$ 180$^\circ$) at elevated pressures is called pressure-induced hydrogen bond symmetrization. This transition can be understood conceptually as the change in the behavior of hydrogen atoms as the potential surface between oxygen atoms decreases from a double minima to a single minimum with increased pressure \citep{tsuchiya2022}. In ice, hydrogen bond symmetrization can be described as a change in the state of hydrogen in the body-centered cubic sublattice of oxygen \citep{hemley1987}. Previous studies have shown that in the ice VII to ice X transition, hydrogen changes from static (ice VII) to dynamically disordered (ice VII$'$ and ice X$'$) and then to symmetric hydrogen bonding states (ice X). In the dynamically disordered state of ice, hydrogen atoms move back and forth between two minima which exist between oxygen atoms due to thermal and quantum vibrational effects (i.e. tunneling effects) \citep{benoit1998, morrone2009}. However, there is no consensus regarding the pressures at which these phase transitions occur or what physicochemical changes mark these transitions \citep{aoki1996, goncharov1996, sugimura2008,caracas2008}.

The existence of dynamically disordered ice has been established through ab initio molecular dynamics simulations (AIMD) \citep{benoit2002}, ab initio lattice dynamics calculations \citep{caracas2008}, and high-pressure experimental studies of ice VII \citep{sugimura2008}. Sugimura et al. \cite{sugimura2008} reported that the compressibility of high-pressure ice changes at 40 and 60 GPa, corresponding to a transition to dynamically disordered ice VII (ice VII$'$) at 40 GPa, followed by a transition to the dynamically disordered ice X (ice X$'$) structure at 60 GPa. That study \cite{sugimura2008}, also reported that the experimentally obtained volumes of the dynamically disordered phases were noticeably smaller than the volumes of static disordered ice VII, which had been calculated from ab initio calculations. Recently, the equation of state and vibrational properties of the high-pressure phases of ice have been reported using ab initio centroid molecular dynamics (CMD) calculations that can properly handle nuclear quantum effects (NQEs) \citep{ikeda2018}. Although, the reduction of volume in the dynamically disordered phases is smaller than the previously reported experimental values, the resulting equation of state does show a volume reduction in the range of 20-60 GPa when compared to conventional AIMD calculations that treat nuclei as classical particles.

If this change of ice to a dynamically disordered state under high pressure affects the equation of state, then it is expected that changes in elastic properties will also appear. The sound velocities of polycrystalline samples of high-pressure ice have been investigated by Brillouin scattering at room temperature and pressures up to 100 GPa \citep{ahart2011}. However, Ahart et al. \citep{ahart2011} reported that the elastic constants of ice increase monotonically and without discontinuities with increased pressure in the pressure range they investigated. More recently, similar conclusions have been reported for the elastic constants of single-crystal ice by Brillouin scattering \citep{zhang2019}. In this study, the elastic constants were smaller than those measured for polycrystalline samples at similar pressures, but were still reported to evolve smoothly with increasing pressure. Conversely, it has been claimed that the elastic constants of ice containing a 0.9 mol.\% of sodium chloride, show a clear softening in the elasticity at a pressure range of 42-54 GPa corresponding to the dynamic disordered phase of ice \citep{shi2021}. 

In light of this disagreement regarding the experimentally determined elastic properties of ice, this study investigated the temperature and quantum effects of hydrogen on the elastic properties of ice. As previously described, earlier CMD calculations have shown that NQEs affect the curvature of the equation of state in dynamically disordered states, and that the elastic properties are expected to be affected accordingly. Therefore, in this study, we performed first-principles path integral molecular dynamics (PIMD) calculations \citep{marx1994, marx1996, tuckerman1996, shiga2001} to determine the elastic properties of ice at room temperature and high pressures, taking into account NQEs, to determine the elasticity of high pressure phase of ice and explore to what extent the NQE influences the elastic properties of the dynamically disordered phase of ice.

\section{\label{sec:level2} Methods}
In this study, ab initio path-integral molecular dynamics (PIMD) simulations based on the Born-Oppenheimer approximation were performed using the PIMD software package (https://ccse.jaea.go.jp/software/PIMD/index.en.html) \citep{shiga2001} along with the open-source electronic structure calculation package Quantum-espresso \citep{giannozzi2009}, which is based on density functional theory \citep{hohenberg1964, kohn1965}. In order to properly evaluate the quantum effects of hydrogen nuclei in the high-pressure ice phase, the equation of state and elastic constants were calculated using PIMD at 300 K and high pressures, replicating the pressure-temperature conditions of most experiments on these phases. The generalized gradient approximation in the Perdew-Burke-Ernzerhof form was used for the exchange correlation functionals, and norm-conserving pseudopotentials were used for the ionic inner-shell potentials of hydrogen and oxygen atoms \citep{troullier1991}. The kinetic energy cutoff in the plane-wave expansion of the wavefunctions was set to 80 Ry. These pseudopotentials have been extensively tested in the calculations of several hydrous minerals and ice \citep[e.g.,][]{tsuchiya2009,tsuchiya2017}. The sampling of the Brillouin zone was limited to the $\Gamma$ point. The calculation of ice VII was modeled using a supercell (3$\times$3$\times$3 unit cells) containing 54 H$_2$O molecules. We performed calculations treating each of 162 atoms as the 10 and 32 beads in the simulation cells. Our PIMD simulations were within canonical (NVT) ensemble where the number of atoms (N) and the cell volume (V) were set to a constant temperature (T), which was controlled by the massive Nos\'{e}-Hoover chains method. The time step $\Delta t$ was set to 0.1 fs, and after equilibrating the system for about 1 ps, the trajectories were collected at least for 1 ps to accumulate statistics. To enable comparison with PIMD, ab initio molecular dynamics (AIMD) calculations were also performed on the same system and under same calculation conditions except for the number of beads.

In this study, it was important to estimate the effect of NQEs on pressure and stress. Because of the spatial extent of the hydrogen nuclei, the quantum effects of the nuclei on the pressure are also expected to be significant. To properly incorporate this effect, a virial correction was performed, which adds the product of the distance from the centroid of the beads and the force applied to each bead \citep{martyna1999}.
The isothermal elastic constants C$_{ij}$s were calculated from the linear relationship of stress-strain by applying a strain of $\pm$0.01 and time-averaging the induced stresses. We confirmed that a linear relation was ensured for this strain range using a previously established methodology \citep{tsuchiya2017}.

\begin{figure}[b]
\includegraphics[width=8cm]{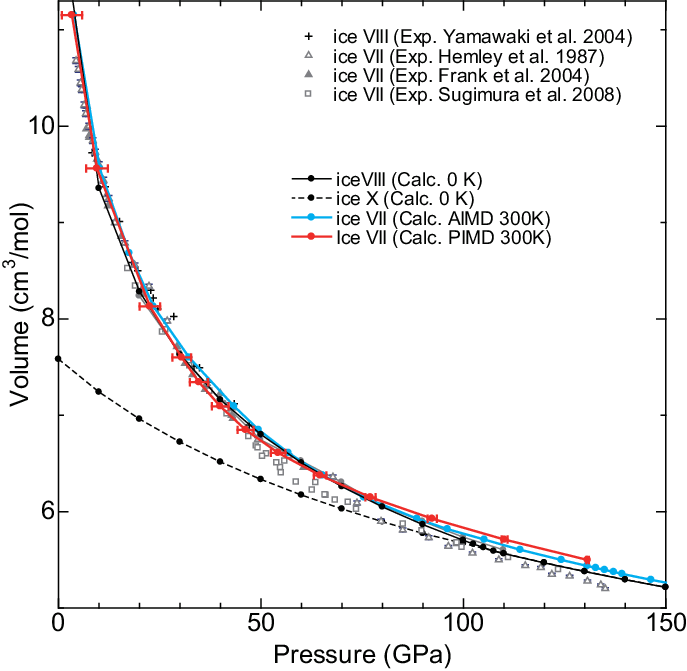}
\caption{\label{eos} The pressure (P) - volume (V) relationship of ice VII under pressure. The black solid and dashed lines indicate the P-V curves of ice VIII (asymmetric hydrogen bond with ordered hydrogen positions) and ice X (symmetric hydrogen bonds) at static 0 K condition, respectively. The red and blue lines indicate the P-V relationships of ice VII at 300 K obtained by PIMD and AIMD simulations, respectively. The volumes determined by PIMD simulation are smaller than those obtained by AIMD simulation at 30-60 GPa. The experimental data of \citet{sugimura2008} and \citet{hemley1987} were reanalyzed using the recent pressure scales of \citet{dorogokupets2007} and \citet{dewaele2004}, respectively.}
\end{figure}

\begin{figure*}[b]
\includegraphics[width=\textwidth]{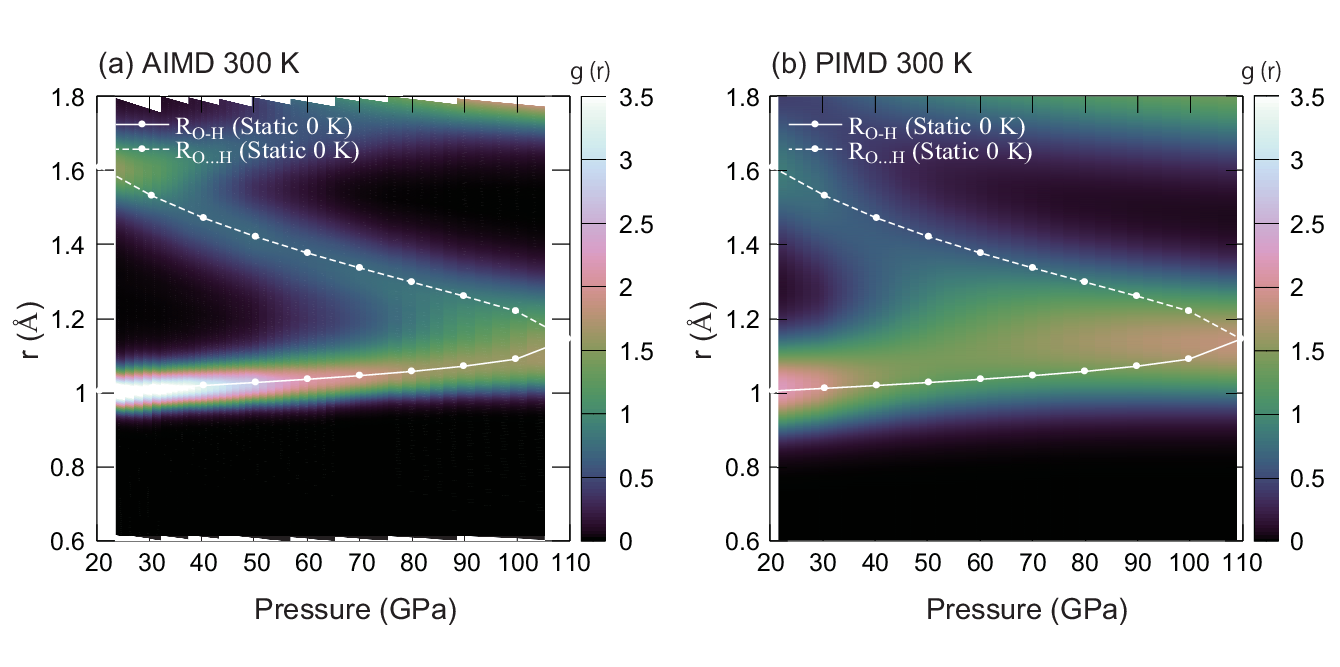}
\caption{\label{rdf} The radial distribution function (RDF) under pressure. The white solid and dashed lines are the OH and O$\cdots$H distances for the ice VIII phase, respectively, at static 0 K conditions by structural optimization that does not take into account the quantum effects of nuclei.
}
\end{figure*}

\begin{figure}[b]
\includegraphics{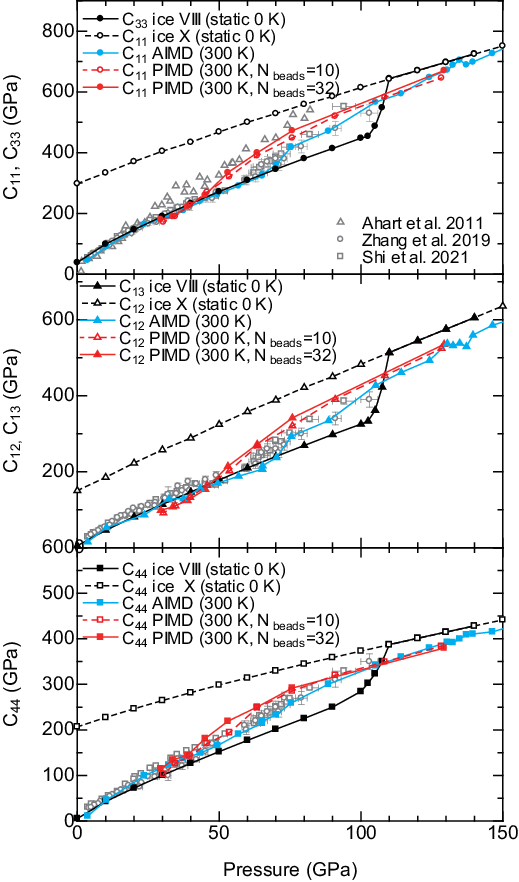}
\caption{\label{cij} The elastic constants of ice VII/VIII and ice X under pressure. The black full and dashed lines indicate the elastic constants of ice VIII and X at static 0 K condition, respectively. The elastic constants of the ice X phase are calculated by distorting them under the constraint that the original space group is maintained. The red and blue lines indicate the C$_{ij}$ of ice VII at 300 K determined by PIMD (N$_{beads}$=32) and AIMD simulations, respectively. Purple lines indicate PIMD simulation with smaller number of beads (N$_{beads}$=10). The open symbols are previous Brillouin scattering measurements \citep{ahart2011, zhang2019, shi2021}.}
\end{figure}
\section{\label{sec:level3}Results\protect}
First, we compare the volumes of ice VII under pressure at 300 K calculated using PIMD and AIMD (Figure \ref{eos}). The volumes obtained by PIMD shows a slight decrease between 20-70 GPa compared to those of AIMD. This result is similar to the equation of state of ice VII obtained by CMD \citep{ikeda2018}. The P-V relationship obtained by PIMD agrees well with previously published experimental results up to 50 GPa, but is larger than the experimental results at higher pressures \citep{sugimura2008, hemley1987}. This trend is also true above 100 GPa, where the hydrogen bonds in the ice are expected to be already symmetrized. Moreover, these experimental results at room temperature (300 K) have a smaller volume than the ab initio results at static 0 K. These discrepancies between experiments and calculations may be partially attributed to differences in the pressure calibration used in the experiments \citep{shi2021}, or pressure gradients within the samples. In particular, the ruby pressure standard used by \citet{hemley1987} produces significant errors under non-hydrostatic conditions, particularly at pressures exceeding $\sim$50 GPa, as the R$_1$ fluorescence line broadens under these conditions.

To investigate the influence of NQEs on the hydrogen bonding properties of ice VII under pressure, the radial distribution function (RDF) of ice VII was obtained using both PIMD and AIMD results, as shown in Figure \ref{rdf}. The RDF obtained by PIMD exhibits a higher distribution between the covalent ($R_{O-H} \sim 1$ \AA) and hydrogen bond distances ($R_{O\cdots H}\sim 1.3-1.6$ \AA) at lower pressures than that obtained by AIMD. This suggests that partial hydrogen bond symmetrization occurs at lower pressures due to quantum effects of hydrogen nuclei. 

Figure \ref{cij} shows the elastic constants of ice VII/VIII and ice X under pressure. The solid and dashed black lines are the elastic constants of ice VIII and X previously reported at static 0 K conditions \citep{tsuchiya2017}. The elastic constants of ice VIII (solid black lines), which has ordered hydrogen positions, increases steeply at $\sim$100 GPa due to the symmetrization of the hydrogen bonds. The elastic constants of metastable ice X with symmetric hydrogen bonds (dashed black lines) below $\sim$100 GPa are significantly larger than those of ice VIII. The main difference between their structures is whether the hydrogen bonds are symmetrized or not. Thus, even partial hydrogen bond symmetrization at finite temperature conditions in ice VII, as indicated in RDF (Figure 2), is expected to contribute to an increase in the elastic constants.

In order to explore the difference in the elastic constants between the dynamically disordered and statically disordered states of hydrogen, the pressure dependence of elastic constants of various hydrogen disordered structures were calculated at static 0 K (Figure S1). Our findings show that the elastic constants of those structures share almost the same values, and that hydrogen bond symmetrization occurred around 100 GPa. These results suggest that the disordered structures themselves do not contribute to the increase in the elastic constants, but rather that the increase is due to the instantaneous partial hydrogen-bond symmetrization in the dynamically disordered state.

Finally, we compare the elastic constants from AIMD, PIMD, and Brillouin scattering experiments\citep{ahart2011, zhang2019}. All of AIMD, PIMD, and experimental values show almost parallel changes to the static 0 K elastic constants at low pressure conditions, which may be explained only by differences in temperature conditions in the static disordered ice VII phase. However, as pressure is increased, the PIMD, the experimental (single crystal), and the AIMD results begin to notably deviate from the static 0 K values at 40, 60, and 70 GPa, respectively. These increases of the elastic constants are likely caused by the dynamic behavior of hydrogen atoms. The comparison of AIMD and PIMD results indicates that NQE contributes significantly to the increase in elastic constants above 40 GPa and by as much as 20\% at around 70 GPa at 300 K condition. Comparing the pressure change in the elastic constants at static 0 K with the experimental elastic constants, the change is not smooth or softening, as has been suggested, but rather hardening. The pressure at which this discontinuous increase in elastic constants occurs is different between the Brillouin scattering experiment and PIMD results. Possible reasons for this discrepancy could be due to the approximation of the exchange correlation functional in the calculations, or because the Brillouin scattering peak is hidden by the diamond peak at about 50 GPa, making it difficult to measure.

\section{\label{sec:level4}Conclusions\protect}
In conclusion, we show that the elastic constants of ice increase non-linearly due to the partial hydrogen symmetrization associated with the change of state of hydrogen in ice phase VII to a dynamically disordered state. This increase is attributed to temperature and the quantum nature of the nuclei. Nuclear quantum effects (NQE) contribute to the increase in elastic constants by up to 20\% at room temperature and high pressure. 

\begin{table*}[b]
\caption{\label{table1}
The high pressure elastic constants, and the bulk (B) and shear (G) moduli (Hill average) of ice VII at 300 K calculated by PIMD method (N$_{beads}$=32).}
\begin{ruledtabular}
\begin{tabular}{cccccccc}
 P (GPa) &V (cm$^3$/mol)& C$_{11}$  &  C$_{12}$  & C$_{44}$ & B  & G  \\
\hline
30.4 & 7.600 & 179.84 & 98.13 & 114.62 & 125.37 & 75.83  \\
34.7 & 7.343 & 189.83 & 107.63 & 133.94 & 135.03 & 83.58 \\
39.9 & 7.093 & 223.05 & 123.73 & 144.97 & 156.84 & 94.43 \\
46.2 & 6.849 & 260.68 & 153.85 & 181.22 & 189.46 & 111.35 \\
54.2 & 6.610 & 334.55 & 214.32 & 219.19& 254.40 & 131.03 \\
64.6 & 6.377 & 399.59 & 272.04 & 250.22 & 314.56 & 145.49 \\
77.0 & 6.150 & 472.06 & 341.31 & 291.47 &  385.15 & 161.64 \\
130.7 & 5.498 & 670.62 & 535.50 & 379.87 & 580.54 & 194.14 \\
\end{tabular}
\end{ruledtabular}
\end{table*}

\begin{acknowledgments}
This work was supported by JSPS KAKENHI Grant Number JP20K04043, JP20K04126, JP23K04670 and JP23H01273. M.S. was supported by from 'Hydrogenomics' of Grant-in-Aid for Scientific Research on Innovative Areas, MEXT, Japan, JSPS KAKENHI (JP18H05519, JP21H01603), and the JAEA supercomputer project.
\end{acknowledgments}

\bibliography{tsuchiya_pimd}

\end{document}


\maketitle

\section*{Elastic constants of static disordered phase of ice VII at 0 K}
To investigate the effect of static hydrogen disorder in the ice VII phase on the elasticity, elastic constants were calculated for various hydrogen configurations in the ice VII supercell. In 2$\times$2$\times$2 supercell (16 H$_2$O) of ice VII. That supercell is made of two interpenetrating Ic lattices, each of which consists of eight H$_2$O molecules, allowing 80 hydrogen configurations to satisfy the Ice rule. Therefore, there are 90$\times$90$=$8100 possible hydrogen configurations. However, some of them have the same structure, which can be eliminated to 52 irreducible configurations (one of which is the same hydrogen configuration as in the ice VIII phase).
\begin{figure}
\includegraphics[width=10cm]{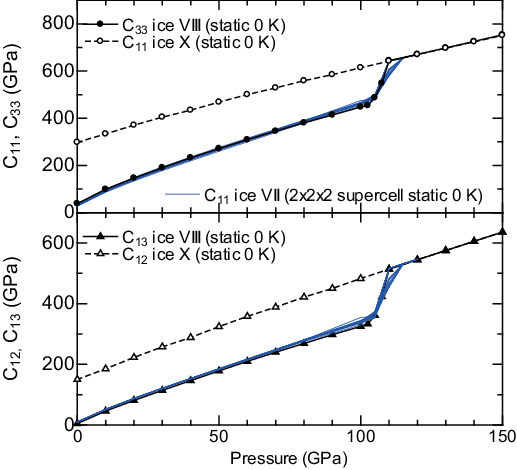}
\caption{Supplementary figure.\label{fig:sup}}
\end{figure}
We calculated the elastic constants of ice VII with these 52 hydrogen configurations at static 0 K conditions. Figure S1 shows the pressure dependence of the elastic constants of ice phase VII with these 52 different hydrogen configurations (full blue lines). These results are almost identical and overlap each other. This is because disordered states with asymmetric hydrogen bonds do not themselves cause an increase in the elastic constants. 




\section*{ab initio molecular dynamics (AIMD) calculation of ice VII under pressure}

\begin{table*}[b]
\caption{\label{tab:table3}
The high pressure elastic constants and moduli (Hill average) of ice VII at 300 K calculated by AIMD method.}
\begin{tabular}{cccccccc}
P (GPa) &V (cm$^3$/mol)& C$_{11}$  &  C$_{12}$  & C$_{44}$ & B  & G  \\
\hline
  3.7	& 11.152 & 46.0 &  15.1 &  11.1 &  25.4	&  12.6 \\
 10.3 &  9.562 & 91.1 &  52.8 &  46.1 &  65.5 &  32.4 \\
 23.6 &	8.130 & 164.1 &  86.2 &  99.4 & 112.2 &  68.3 \\
 32.2 &	7.600 & 190.1 & 127.9 & 123.3 & 148.6 &  71.4 \\
 43.4 &	7.093 & 239.7 & 155.6 & 149.1 & 183.6 &  90.1 \\
 49.5 &	6.849 & 263.0 & 169.7 & 165.8 & 200.8 & 100.1 \\
 56.9 &  6.610 & 290.6 & 188.2 & 191.0 & 222.3 & 113.2 \\
 65.2 &	6.380 & 323.5 & 206.4 & 214.5 & 245.4 & 128.0 \\
 75.4 &  6.150 & 391.9 & 266.6 & 254.0 & 308.3 & 145.9 \\
 88.6 &  5.927 & 473.8 & 342.3 & 297.0 & 386.1 & 164.0 \\
105.2 &  5.710 & 566.5 & 427.3 & 340.9 & 473.7 & 182.8 \\
114.0 &  5.603 & 627.3 & 489.7 & 366.2 & 535.5 & 190.7 \\
124.2 &  5.498 & 670.4 & 520.9 & 376.6 & 570.7 & 199.9 \\
130.6 &  5.436 & 670.8 & 517.7 & 389.6 & 568.7 & 206.1 \\
132.7 &  5.415 & 674.6 & 538.6 & 397.6 & 583.9 & 200.5 \\
134.9 &  5.395 & 659.3 & 491.8 & 397.4 & 547.6 & 215.5 \\
137.1 &  5.374 & 687.8 & 542.0 & 401.2 & 590.6 & 206.5 \\
139.2 &  5.354 & 700.3 & 561.4 & 402.0 & 607.7 & 203.4 \\
146.4 &  5.292 & 726.6 & 585.8 & 415.3 & 632.7 & 208.8 \\
159.0 &  5.191 & 778.4 & 614.0 & 438.1 & 668.8 & 228.0 \\
\hline
\end{tabular}
\end{table*}


